\documentclass[twoside]{article}
\usepackage{amsmath}
\usepackage{amssymb}
\usepackage{latexsym}
\usepackage{revsymb}
\usepackage{epsfig}
\usepackage{graphicx}

\newtheorem{defi}{Definition}

\newtheorem{thm}[defi]{Theorem}
\newtheorem{cor}[defi]{Corollary}
\newtheorem{rem}[defi]{Remark}
\newtheorem{prop}[defi]{Proposition}

\newtheorem{conj}[defi]{Conjecture}

\newcommand{\qed}{\hfill $\Box$}
\newcommand{\tr}{{\operatorname{Tr}}}
\newcommand{\id}{{\operatorname{id}}}

\newcommand{\bra}[1]{{\langle{#1}|}}
\newcommand{\ket}[1]{{|{#1}\rangle}}
\newcommand{\ketbra}[1]{{\ket{#1}\!\bra{#1}}}
\newcommand{\C}{{\mathbb{C}}}
\newcommand{\R}{{\mathbb{R}}}

\newlength{\blank}
\newlength{\equalsign}
\newenvironment{beweis}[1][{\hspace{-\blank}}]{{\noindent\emph{Proof~{#1}.\ }}}{\hfill $\Box$\vskip 0.5\baselineskip}

\begin{document}

\title{\sf \protect{\vspace{-0.5cm}}A new inequality for the von Neumann entropy}
\author{\small Noah Linden\qquad and\qquad Andreas Winter\protect\vspace{-0.09cm}\protect\\
        \small {\tt n.linden@bristol.ac.uk}\qquad
               {\tt a.j.winter@bris.ac.uk}\protect\vspace{-0.09cm}\protect\\
        \small Department of Mathematics, University of Bristol,\protect\vspace{-0.09cm}\protect\\
        \small University Walk, Bristol BS8 1TW, U.K.}
\date{\small (14th June 2004)}

\maketitle

\begin{abstract}
Strong subadditivity of von Neumann entropy, proved in 1973 by
Lieb and Ruskai, is a cornerstone of quantum coding theory. All
other known inequalities for entropies of quantum systems may be
derived from it. Here we prove a new inequality for the von
Neumann
  entropy which we prove is independent of strong subadditivity:
  it is an inequality which is true for any four party quantum
  state,
  provided that it satisfies three linear relations (constraints) on the entropies of certain
  reduced states.
\end{abstract}

\section{Introduction}
\label{sec:intro} Entropy is a key concept both in classical and
in quantum information theory: Shannon's source and channel coding
and half a century of work~\cite{shannon} have exhibited a vast
range of operational coding problems whose solution can be
expressed most naturally by (Shannon-Gibbs-Boltzmann) entropies of
random variables:
$$H(X) = - \sum_x \Pr\{X=x\}\log_2 \Pr\{X=x\}.$$
\par
Quantum information theory~\cite{QIT} allows a wealth of new
information processing possibilities, with the von Neumann
(quantum) entropy playing a role analogous to Shannon's
(classical) entropy in classical information theory: for a density
operator $\rho$, the von Neumann entropy is
$$S(\rho) = - \tr \rho\log_2 \rho.$$
\par
Although the two entropy functionals
exhibit similarities, they have many decidedly different properties.
These properties however, because of the intimate relation of the
entropy to operational properties of (classical and quantum)
information, ultimately express statements about the
``nature of information''. Furthermore, they are indispensable technical
tools in proving the information theoretic optimality
of constructions: most importantly, there are \emph{inequalities}
governing the relative magnitude of entropies, conditional
entropies and mutual informations.
\par
In the quantum case, there is essentially only one known inequality (all others being
derivable from it): strong subadditivity.
Proved by Lieb and Ruskai~\cite{Lieb:Ruskai} in 1973, it
is the key result on which virtually every nontrivial
quantum coding theorem relies.
\par
We prove here a new inequality for the von Neumann entropy,
which we show cannot be derived from the known ones: it is a
\emph{constrained inequality} in that it is not true in general
but only for states satisfying three particular linear constraints
on their entropies.
\par
One starting point for our work was the desire to understand
properties of quantum entropy. We were also motivated by
investigations of multi-party entanglement
in~\cite{Magnificent:7}: there, entropy values were found
which are allowed by strong subadditivity but for which the
authors could not find quantum states. This led to the conjecture
that it is impossible to realise those values by a quantum state,
which we indeed prove here.
\par
The structure of our paper is as follows: the next section
reviews the well-established convexity framework for (linear)
information inequalities. In section~\ref{sec:new} we state and
prove our result, while in section~\ref{sec:why-new} we explain
why it does not follow from the standard inequalities.  In
section~\ref{sec:alternative} we present a number of alternative
forms of our inequality. We close in section~\ref{sec:disc} with a
discussion and a conjectured non-constrained inequality.

\section{Linear inequalities}
\label{sec:linear}
Pippenger~\cite{pippenger:H} initiated the programme of determining all
(linear) inequalities satisfied by the classical entropy functional $H$.
This question was based on the realisation of two facts
(see Yeung's work~\cite{yeung:framework}):
first, that in information theoretic
applications, the properties one uses about the entropy to bound information
quantities seem always to be
\begin{enumerate}
  \item Nonnegativity of entropy $H(X)$.
  \item Nonnegativity of conditional entropy $H(X|Y)=H(XY)-H(Y)$.
  \item Nonnegativity of mutual information $I(X;Y)=H(X)+H(Y)-H(XY)$
    \item Nonnegativity of conditional mutual information $I(X;Z|Y)=H(XY)+H(YZ)-H(XYZ)-H(Y)$,
\end{enumerate}
for random variables $X,Y,Z$, the so-called basic inequalities.
\par
Second, that for every number $n$ of random variables, the points in $\R^{2^n-1}$ given by
the entropies of all possible subsets of the random variables,
$$\bigl\{ (H(X_S))_{\emptyset \neq S \subset \{1,\ldots,n\}}:
                                    X_1,\ldots,X_n\ {\rm random\ variables} \bigr\}$$
(where all random variables are assumed to be discrete and indeed finite range)
form ``almost'' a convex cone (i.e., closed under nonnegative linear combinations)
in the positive orthant: one only needs to go to the
topological closure, denoted $\overline\Gamma^*_n$, and called
the \emph{(classical) entropy cone}.
\par
Surprisingly, the classical entropy cone can be strictly smaller than the cone cut out by
the basic inequalities for all subsets of random variables
(which we will call ``Shannon cone'' $\Gamma_n$):
while the two cones coincide for $n\leq 3$, they differ
for $n=4$. Indeed, as Yeung and Zhang have shown, there are further inequalities
satisfied by the entropy cone which are not dependent on the basic inequalities;
i.e., they are violated by points in the Shannon cone.
\par
Pippenger~\cite{pippenger:S} observed that a similar situation occurs in the
quantum case: with an underlying state multipartite state $\rho$,
denote the entropy of its restriction to subsystems $A,\ldots$,
or groups of subsystems $AB,\ldots$, by $S(A)$, $S(AB)$, etc.
Then, there is a ``von Neumann'' cone $\Sigma_n$,
defined by the basic inequalities
\begin{enumerate}
  \item Nonnegativity of entropy $S(A)$.
  \item Nonnegativity of the quantity $S(A|B)+S(A)=S(AB)-S(B)+S(A)$
    (this is known as the Araki-Lieb inequality~\cite{Araki:Lieb})
  \item
    Nonnegativity of $S(C|A)+S(C|B)=S(CA)+S(CB)-S(A)-S(B)$ (this replaces nonnegativity
    of the conditional entropy, and is called ``weak monotonicity'').
  \item Nonnegativity of quantum mutual information $I(A;B)=S(A)+S(B)-S(AB)$
  \item Nonnegativity of
    quantum conditional mutual information $I(A;C|B)=S(AB)+S(BC)-S(ABC)-S(B)$.
\end{enumerate}
(Note that the names of the quantities are given based on straightforward
analogy, with no operational significance implied at this point.)
\par
The latter two are simply subadditivity and strong
subadditivity~\cite{Lieb:Ruskai} of the quantum entropy.
The properties 2) and 3) above, can actually be derived from 4) and 5) (and vice
versa) by viewing the state as the restriction of a pure state
on the given parties plus one, and the fact that for a pure state,
the entropy of a subset of the parties equals the entropy of the
complementary set. This is  a consequence of linear
algebra, namely the Schmidt decomposition of bipartite pure states,
whose coefficients are the eigenvalues of both reduced states
(for a more detailed discussion, see~\cite{pippenger:S}).
Note also that choosing trivial $B$ (i.e., with Hilbert space $\C$)
reduces weak monotonicity to the Araki-Lieb inequality, and
conditional mutual information to mutual information
(compare the classical case).  Thus all non-trivial inequalities may be derived
from strong subadditivity.
\par
And there is the cone of the closure of all points realized by
entropies of the $2^n-1$ nontrivial marginals,
the (quantum) entropy cone $\overline\Sigma^*_n$,
of states on tensor products of $n$ finite quantum systems
(that it is indeed a cone is proved in the same way as for the classical
case~\cite{pippenger:S}).
\par
The faces of the cone $\Sigma_n$ are given by certain entropies
being zero (which means that the corresponding subsystem is in a pure state),
certain mutual informations being zero (which means that certain pairs of
subsystems are in a product state), certain conditional mutual
informations being zero, etc.
The latter is fully analogous to the classical
case of a Markov chain, where $A$ and $C$ are independent conditional
on $B$ (which we call the ``pivot'' of the chain), as
explained in~\cite{SSA-saturation}.
\par\medskip
All this raises the following natural question: are there any further
linear inequalities for the quantum entropy than those above?
To be precise: is $\overline\Sigma^*_n \subsetneq \Sigma_n$, and if
so, can we find a hyperplane intersecting the interior or $\Sigma_n$
but having $\overline\Sigma^*_n$ entirely in one halfspace?

\section{The new inequality}
\label{sec:new}
Our main result is the following theorem, which gives an answer to
the question in its first form, and provides evidence for a
positive answer to the second.
\begin{thm}
  \label{thm:cei}
  Let $\rho^{ABCD}$ be a state of a quadripartite quantum system, such that
  strong subadditivity is saturated for the three triples $ABC$, $CAB$
  and $ADB$ (pivot always in the middle). Then,
  $$I(C;D) \geq I(C;AB).$$
\end{thm}
\begin{figure}[h]
  \label{fig:cei-diagram}
  \centering
  \includegraphics{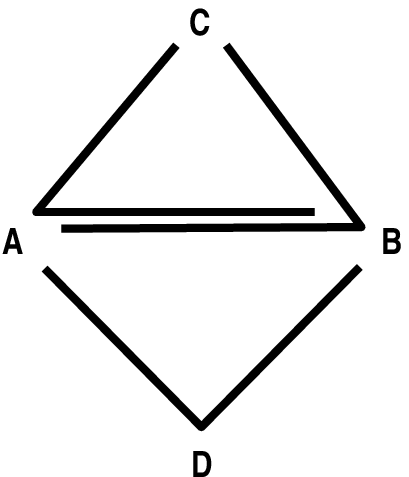}
  \qquad
  \begin{minipage}{7.2cm}
    \vspace{-2.6cm}
    {\small
      The angles in the figure represent the strong subadditivity constraints
      which are saturated in the conditions of theorem~\protect\ref{thm:cei}
      (i.e. strong subadditivity is
      saturated for the triples $ABC$, $CAB$, $ADB$; pivot always in the middle).
      Theorem~\protect\ref{thm:cei} then states that under these conditions,
      the correlation between $C$ and $D$ is not smaller than that between $C$ and $AB$.
    }
  \end{minipage}
\end{figure}
\begin{beweis}
  The proof relies heavily on the recent characterisation of
  states which saturate strong subadditivity~\cite{SSA-saturation},
  which is stated below as proposition~\ref{prop:SSA-eq}.
  \par
  First of all, since we have strong subadditivity saturated for $CAB$,
  $$\rho^{CAB} = \bigoplus_{i} p_i\, \rho_i^{C a_i^L} \otimes \rho_i^{a_i^R B},$$
  by proposition~\ref{prop:SSA-eq}. To this we apply proposition~\ref{prop:SSA-eq}
  once more, for the triple $ABC$: there exists the recovery map $R_{B\rightarrow C'}$
  (we duplicate $C$, and attach a prime, to distinguish the
  two incarnations of $C$) mapping $\rho^{AB}$ to $\rho^{ABC'}$.
  It maps the above $\rho^{CAB}$ to
  $$\rho^{CABC'} = \bigoplus_{ij} p_i p_{j|i}\,
                         \rho_i^{C a_i^L} \otimes \rho_{ij}^{a_i^R b_j^L}
                                          \otimes \rho_j^{b_j^R C'}.$$
  (Notice that the states on the far right, by the structure of $R_{B\rightarrow C'}$
  can only depend on $j$, the Hilbert space sector measured by the map, as described
  in proposition~\ref{prop:SSA-eq} below.)
  \par
  But the two states obtained by tracing out $C$ and $C'$, respectively (and
  identifying $C$ with $C'$ again), must coincide:
  \begin{align*}
    \rho^{CAB} &= {\textstyle\bigoplus_{ij}}\  p_{ij}\,
                     \rho_i^{C a_i^L} \otimes \rho_{ij}^{a_i^R b_j^L} \otimes \rho_j^{b_j^R} \\
    \parallel\phantom{==} & \\
    \rho^{ABC} &= {\textstyle\bigoplus_{ij}}\  p_{ij}\,
                     \rho_i^{a_i^L} \otimes \rho_{ij}^{a_i^R b_j^L} \otimes \rho_j^{b_j^R C}
  \end{align*}
  Comparing these two, for a given sector labelled $ij$, with $p_{ij}>0$,
  we obtain that both $\rho_i^{C a_i^L}$ and $\rho_j^{b_j^R C}$ are actually
  product states:
  \begin{equation}\begin{split}
    \label{k-sectors}
    \rho_i^{C a_i^L} &= \rho_j^C \otimes \rho_i^{a_i^L}, \\
    \rho_j^{b_j^R C} &= \rho_j^{b_j^R} \otimes \rho_i^C.
  \end{split}\end{equation}
  That the right hand sides contain both $i$ and $j$ (whereas the left hand sides
  mention only $i$ and only $j$, respectively) is no error, but in fact the main point:
  it means that for actually occurring $ij$, i.e., $p_{ij}>0$, the
  state of $C$ belonging to this sector depends only on $i$ \emph{and}
  only on $j$ --- in other words, it must be a common function of $i$
  and $j$: $\rho_k^C$, with $k=f(i)=g(j)$, with certain (deterministic)
  functions $f$ and $g$.
  \par
  We note that if all the $p_{ij}$ are strictly positive, then the only way
  for $\rho^C_k$ to be consistent with eq.~(\ref{k-sectors}) is for it to be constant,
  i.e. independent of $i$ and $j$.
  Situations in which  $\rho^C_k$ can vary with $i$ and $j$ are only possible
  when some of the $p_{ij}$ are zero.  As an illustration, consider a state
  $\rho^{ABC}$ which has $i,j=1,2,3$, and $p_{11}>0,\ p_{22}>0,\ p_{23}>0,\ p_{32}>0$
  and $p_{33}>0$, but $p_{ij}=0$ otherwise; then the non-constant possibility
  $\rho^C_k=\rho_1$ for $i=j=1$ and $\rho^C_k=\rho_2\neq\rho_1$ for
  $(i,j)=(2,2),(2,3),(3,2),(3,3)$ (i.e. $f(1)=g(1)=1,\ f(2)=f(3)=g(2)=g(3)=2$)
  is consistent with eq.~(\ref{k-sectors}).
  \par
  Returning to the general situation,
  with $k$ as described in the previous paragraph but one,
  we can rewrite $\rho^{ABC}$ again,
  $$\rho^{ABC} = \bigoplus_{ij} p_{ij}\,
                   \rho_i^{a_i^L} \otimes \rho_{ij}^{a_i^R b_j^L}
                             \otimes \rho_j^{b_j^R} \otimes \rho_k^C.$$
  In fact, let us introduce quantum registers $K_A$ and $K_B$
  holding $k$ explicitly (of course, by our observation, they will be perfectly
  correlated) --- and note that their content can be extracted locally at $A$
  and $B$, respectively, without disturbing the state $\rho^{ABC}$, by a measurement
  of the orthogonal subspace sector $i$ and $j$, respectively:
  $$\rho^{K_AA BK_B C}
      = \bigoplus_{ij} p_{ij}\, \ketbra{k}^{K_A} \otimes \rho_i^{a_i^L}
                                  \otimes \rho_{ij}^{a_i^R b_j^L} \otimes
                                \rho_j^{b_j^R} \otimes \ketbra{k}^{K_B} \otimes \rho_k^C.$$
  We shall use the following convention: some of the registers are classical
  (such as $K_A$ and $K_B$) inasmuch they come with a distinguished basis,
  and the global state is written as a mixture of states which have the
  classical registers in one of their distinguished basis states.
  These classical registers we identify with random variables, by the
  same name, for example $K_A$ with distribution
  $$\Pr\{K_A=k\}=p_k=\sum_{i:f(i)=k} p_i.$$
  This will allow us to speak about the state in quantum theoretical language,
  and interchangeably about its classical properties in random variable language.
  For example, as random variables, $K_A=K_B$ with probability $1$,
  by our earlier observation.
  \par
  The proof will now be completed by showing two things:
  first, that $I(C;AB)$ equals the Holevo quantity of the ensemble of
  the $\rho_k^C$, which is $I(K_A;C)$;
  second, that $k$ is also ``known at $D$'' by which we mean that there
  is a measurement on $D$ extracting a random variable $K_D$ perfectly
  correlated with $K_A=K_B$.
  \par
  First, the first claim: the system $AB$, by our above
  characterisation, falls into orthogonal sectors $(AB)_k$, labelled by $k$,
  and the state in this sector is some $\sigma_k^{(AB)_k}\otimes\rho_k^C$,
  because it is a convex combination of states
  $\rho_i^{a_i^L} \otimes \rho_{ij}^{a_i^R b_j^L} \otimes \rho_j^{b_j^R} \otimes \rho_k^C$,
  with $ij$ consistent with $k$, so they all have the same state $\rho_k^C$ on $C$.
  Hence, there is a quantum operation extracting $K_A$ from $AB$
  (a coarse-graining of the disturbance-free measurement of $i$ and $j$),
  as well as a reverse, creating $\sigma_k^{(AB)_k}$ in $AB$ from $K_A$.
  By monotonicity of the quantum mutual information, $I(C;AB)=I(K_A;C)$.
  \par
  The second claim is seen as follows: using the third constraint,
  $I(A;B|D)=0$, with the monotonicity of the quantum conditional mutual
  information under the local maps extracting $K_A$ (from $A$)
  and $K_B$ (from $B$),
  gives $I(K_A;K_B|D)=0$. Proposition~\ref{prop:SSA-eq} guarantees the
  existence of a measurement (whose outcome we think of being stored in
  a classical register $K_D$)
  such that conditional on each measurement outcome, $K_A$ and $K_B$
  are in a product state. It is straightforward to check that
  then $I(K_A;K_B|K_D)=0$, and because $K_A$ and $K_B$ are perfectly
  correlated, they must also be perfectly correlated with $K_D$:
  $K_A=K_B=K_D$ with probability $1$, as random variables.
  \par
  These two facts, by monotonicity of the quantum mutual
  information under quantum operations, finally
  yield $I(C;AB) = I(K_A;C) = I(K_D;C) \leq I(C;D)$.
\end{beweis}

\begin{prop}[Hayden, Jozsa, Petz and Winter~\cite{SSA-saturation}]
  \label{prop:SSA-eq}
  A state $\rho^{ABC}$ saturating strong subadditivity at pivot $B$,
  i.e., satisfying $S(AB)+S(BC)=S(ABC)+S(B)$, must have the following form.
  There exists an orthogonal decomposition of
  $B$'s Hilbert space ${\cal H}_B$ into subspaces ${\cal H}_{B_j}$, each of which
  has a natural presentation as tensor product of two Hilbert spaces:
  $${\cal H}_B = \bigoplus_j {\cal H}_{b_j^L} \otimes {\cal H}_{b_j^R},$$
  such that (with states $\rho_j^{Ab_j^L}$ on ${\cal H}_A\otimes{\cal H}_{b_j^L}$
  and $\rho_j^{b_j^R C}$ on ${\cal H}_{b_j^R}\otimes{\cal H}_C$)
  $$\rho^{ABC} = \bigoplus_j p_j\, \rho_j^{Ab_j^L}\otimes \rho_j^{b_j^R C}.$$
  This can be operationally rephrased as follows:
  there is a quantum operation
  $R_{B\rightarrow C}$ from $B$ to $BC$ such that
  $\rho^{ABC}=(\id_A\otimes R_{B\rightarrow C})\rho^{AB}$, which has the
  following form:
  \begin{enumerate}
    \item Perform a projective measurement associated with an orthogonal
      decomposition of $B$ into sectors $B_j$.
    \item Each sector has a tensor product structure $B_j = b_j^L b_j^R$;
      having measured $j$ in step 1,
      the map discards the state on $b_j^R$ and replaces it
      by $\rho_j^{b_j^R C}$ on the composite system $b_j^R C$.
      \qed
  \end{enumerate}
\end{prop}

\begin{rem}
  \label{rem:inequality}
  \rm
  One can easily construct states where our inequality
  is strict, and others where it is tight: a state of the form
  $$\rho^{ABCD} = \bigoplus_j q_j \rho_j^A \otimes \sigma_j^B \otimes \varphi_j^{CD},$$
  with $\varphi_j^{CD}$ being arbitrary and having the marginal states
  $\tau_j$ and $\zeta_j$ on $C$
  and $D$, respectively, will generically have $I(C;D) > I(C;AB)$. If however
  $\varphi_j^{CD} = \tau_j^C \otimes \zeta_j^D$, with mutually
  orthogonal $\tau_j^C$, we have equality.
  \par\medskip
  Also, it is easy to see that no proper subset of our three constraints
  can imply $I(C;D)\geq I(C;AB)$:
  \begin{enumerate}
    \item Saturation of strong subadditivity for $ABC$ and $CAB$: consider
      $$\rho^{ABCD}=\frac{1}{2}\bigl( \ketbra{000}+\ketbra{111} \bigr)^{ABC}
                              \otimes \ketbra{0}^D.$$
      It satisfies these two constraints (and may more),
      but has $I(C;D)=0$ and $I(C;AB)=1$.
    \item Saturation of strong subadditivity for $ABC$ and $ADB$: consider
      $$\rho^{ABCD}=\frac{1}{2}\bigl( \ketbra{000}+\ketbra{111} \bigr)^{BC}
                              \otimes \ketbra{00}^{AD}.$$
      It satisfies these two constraints (and many more),
      but has $I(C;D)=0$ and $I(C;AB)=1$.
  \end{enumerate}
\end{rem}

\begin{rem}
  \label{rem:nontrivial}
  \rm
  It is worth pointing out that not every application of
  proposition~\ref{prop:SSA-eq} along the lines of our proof
  of theorem~\ref{thm:cei} yields a nontrivial result, even though
  it may seem so at first sight: for example, consider a tripartite
  state $\rho^{ABC}$ which saturates strong subadditivity for $ABC$. Then, the
  characterisation of such states implies that $\rho^{AC}$ is
  separable, which is well-known to imply $S(AC)\geq S(C)$.
  Since this inequality is false for general states, have we found a
  new constrained inequality?
  Actually no: it can be checked immediately that in generality,
  $$2S(A|C) + I(A;C|B) = \bigl[ S(A|B)+S(A|C) \bigr] + \bigl[ I(A;B|C) \bigr] \geq 0,$$
  by the basic inequalities (weak monotonicity and strong subadditivity):
  hence, if $I(A;C|B)=0$, then necessarily $S(A|C)\geq 0$.
\end{rem}

\section{Why the constrained inequality is new}
\label{sec:why-new}
In the introduction we have explained already why for three parties there
cannot be an information inequality independent of the basic ones,
as $\overline\Sigma^*_3=\Sigma_3$~\cite{pippenger:S,Magnificent:7}.
Indeed, as one can see from these papers there cannot even be a
constrained inequality, since on each of the $8$ extremal rays
of $\Sigma_3$ there are (nonzero) entropy vectors realised by certain
states.
\par
The four party case is studied in~\cite{Magnificent:7} with particular interest
in the insights to be gained about multi-party entanglement.
 There it is shown that $\Sigma_4$ has 76
extremal rays, which fall naturally into $8$ classes by symmetries
(permutation of the parties).
For $6$ of them~\cite{Magnificent:7} gives states
realising entropy vectors on the rays. The two remaining classes are
represented by the rays spanned by the following vectors
(the first row gives the combinations of subsystems in lexicographic
order; below are their ``entropies''):
\par\medskip
\noindent
\begin{tabular}{l||c|c|c|c|c|c|c|c|c|c|c|c|c|c|c}
            & A & B & C & D & AB & AC & AD & BC & BD & CD & ABC & ABD & ACD & BCD & ABCD \\
  \hline\hline
  \text{I}  & 3 & 3 & 2 & 2 &  4 &  3 &  3 &  3 &  3 &  4 &   4 &   4 &   3 &   3 &    2 \\
  \hline
  \text{II} & 3 & 3 & 3 & 3 &  4 &  4 &  4 &  4 &  4 &  6 &   5 &   5 &   5 &   5 &    2 \\
\end{tabular}
\medskip\\
Clearly, if one could find states realising these vectors (or nonzero
multiples), this would prove $\overline\Sigma^*_4 = \Sigma_4$.
\par
It is readily verified that both these rays satisfy the condition of
theorem~\ref{thm:cei}, but not the conclusion: both vectors given above
have $I(C;D)=S(C)+S(D)-S(CD)=0$ but $I(C;AB)=S(C)+S(AB)-S(ABC)=2>0$.
\begin{cor}
  \label{cor:no-I-II}
  There are no quantum states of finite systems realising entropy vectors
  on the rays I and II above. In fact, in the face of the cone $\Sigma_4$
  described by the three constraint equations of theorem~\ref{thm:cei}, the new
  inequality $I(C;D)\geq I(C;AB)$ cuts off a slice, which contains the
  rays I and II.
  \qed
\end{cor}
In other words, the two entropy vectors satisfy all the basic inequalities,
but by  theorem~\ref{thm:cei} there can be no non-trivial quantum state
with entropy vector in these rays.
\par
Thus, theorem~\ref{thm:cei} cannot be derived from the constraints
in its statement using only the basic inequalities, and so the
new inequality is indeed independent of all previously known
inequalities.

\section{Alternative forms of the inequality}
\label{sec:alternative} We have presented the new inequality in
theorem~\ref{thm:cei} in a form which reflects our way of proving
it. Writing out the mutual informations in terms of entropies, one
notices that some terms cancel, and we arrive at the following
reformulation of our result:
\par\medskip\noindent
{\bf Theorem $\mathbf{1'}\,$}
{\it Let $\rho^{ABCD}$ be a state of a quadripartite quantum
system such that
\begin{align*}
  S(AB) + S(BC) - S(B) - S(ABC) &= 0, \\
  S(CA) + S(AB) - S(A) - S(CAB) &= 0, \\
  S(AD) + S(DB) - S(D) - S(ADB) &= 0.
\end{align*}
Then, $S(ABC)+S(D) \geq S(AB)+S(CD),\text{ i.e., }I(ABC;D)\geq I(AB;CD)$.
\qed}
\par\medskip
We present this reformulation mainly because it may help understanding
and applying the result.
\par
There is another one, however, which is less trivial: we can apply the
purification trick that is used to relate strong subadditivity
and weak monotonicity (see section~\ref{sec:linear}).
\par
In detail, we construct a purification $\Psi^{ABCDE}$ of the
given state $\rho^{ABCD}$ and can apply theorem~\ref{thm:cei} or $1'$
to three situations: strong subadditivity saturated for the triples
$ABC$, $CAB$ and $AEB$; second, for $AEC$, $CAE$ and $ADE$;
third, for $ABE$, $EAB$ and $ADB$. If we then systematically
eliminate all entropies involving $E$ by substituting the complementary
group, we get the following statements:
\par\medskip\noindent
{\bf Theorem $\mathbf{1''}\,$} {\it Let $\rho^{ABCD}$ be a state
of a quadripartite quantum system. Consider the following three
properties this state could have:
\begin{equation*}
\begin{array}{lllllr}
  \phantom{===**} &
  I(A;C|B)       &= I(B;C|A)       &= I(A;B|CD)      &= 0  & \qquad\qquad (1) \\
  \phantom{===**} &
  S(C|A)+S(C|BD) &= I(A;C|BD)      &= S(A|D)+S(A|BC) &= 0  & \qquad\qquad (2) \\
  \phantom{===**} &
  S(B|A)+S(B|CD) &= S(A|B)+S(A|CD) &= I(A;B|D)       &= 0. & \qquad\qquad (3)
\end{array}
\end{equation*}
Then,
\begin{align*}
  (1) &\Longrightarrow  S(C|AB) + S(C|ABD) \geq 0, \\
  (2) &\Longrightarrow  S(C|D) + S(C|BD)   \leq 0, \\
  (3) &\Longrightarrow  S(D)+S(CD) \geq S(AB)+S(ABC).
\end{align*}
\qed }

\section{Discussion}
\label{sec:disc}
Although we believe that
the discovery of a new constrained information inequality is
interesting in itself, our theorem~\ref{thm:cei} is not enough to conclude
$\overline\Sigma^*_4 \subsetneq \Sigma_4$ because it may be that there are
states realising entropy vectors arbitrarily close to the points I and II
in the previous section.
Such a possibility could be ruled out by
finding an \emph{unconstrained} inequality satisfied by
$\Sigma^*_4$ but violated by points on the rays I and II.
Note that indeed for $n=3$, in both the quantum and
classical version of the question,
the set of entropic vectors is not closed, so is not identical
to the entropy cone $\overline\Sigma^*_3$, $\overline\Gamma^*_3$.
On the other hand, it is still the case that the extremal rays are
indeed populated by distributions/states.
\par
We may remark that in the classical variant of the question, Yeung and Zhang also
at first only found a constrained inequality~\cite{yeung:zhang:CEI},
and only somewhat later their unconstrained inequality
in~\cite{yeung:zhang:UCEI}, whose proof indeed uses ideas from
constrained inequalities.
\par
We think, however, that our result provides some evidence towards the
existence of such an inequality for the quantum entropy cone; in fact,
we believe that it way well be possible to prove an inequality
ruling out the approximability
of I and II, based on the following: in~\cite{SSA-saturation}, it is
conjectured that there is a robust version of that paper's main theorem
--- characterising the states that come close to saturating strong
subadditivity. It seems likely that with such a theorem one could
perform an approximation version of the proof of theorem~\ref{thm:cei},
and conclude a new ``constrained'' inequality if the three constraint
equations of theorem~\ref{thm:cei} are only almost satisfied.
In other words, there would be a trade-off between the degree
by which $I(A;C|B),\ I(C;B|A),\ I(A;B|D)$ are nonzero, and
the negativity of $I(C;D)-I(C;AB)$.
\par
This rationalises the following conjecture, with which we close
the paper:
\begin{conj}
  \label{conj:daring}
  There exist positive constants $\kappa_1,\ \kappa_2$ and $\kappa_3$,
  such that for all quadripartite states,
  $$\kappa_1 I(A;C|B) + \kappa_2 I(C;B|A) + \kappa_3 I(A;B|D)
                     + \bigl[ I(C;D)-I(C;AB) \bigr] \geq 0.$$
\end{conj}

\section*{Acknowledgements}
We thank E Maneva, S Massar, S Popescu,
D Roberts, B Schumacher, J A Smolin and A V Thapliyal,
for illuminating discussions on the subjects of this paper over
many years and for allowing us to use their results prior to publication.
We also thank M Christandl and T Osborne for helpful remarks.
\par
Both authors received support from the EU under European Commission
project RESQ (contract IST-2001-37559).

\end{document}